\documentclass[11pt]{article}
\usepackage[utf8]{inputenc}
\usepackage{geometry}
\usepackage{titling}
\usepackage{authblk}
\usepackage{amssymb}
\usepackage{amsmath}
\usepackage{amsthm}
\allowdisplaybreaks[4]
\usepackage{slashed}
\usepackage{graphicx,color,epsfig}
\usepackage{cite}
\usepackage[colorlinks=true,linkcolor=red,citecolor=blue]{hyperref}
\begin{document}
\title{Correlating Gravitational Waves with $W$-boson Mass, FIMP Dark Matter, and Majorana Seesaw Mechanism}
\author{Xuewen Liu}
\author{Shu-Yuan Guo}
\author{Bin Zhu$\footnote{zhubin@mail.nankai.edu.cn}$}
\author{Ying Li$\footnote{liying@ytu.edu.cn}$}
\affil{\it Department of Physics, Yantai University, Yantai 264005, China}
\maketitle
\vspace{0.2cm}

\begin{abstract}
We study a minimal extension of the Standard Model by introducing three right-handed neutrinos and a new scotogenic scalar doublet, in which the mass splittings between neutral and charged components are responsible for the $W$-boson mass newly measured by the CDF collaboration. This model can not only generate non-vanishing Majorana neutrino masses via the interaction of right-handed neutrinos and scotogenic scalars, but also explain the Universe’s missing matter in the form of FIMP dark matter. We also study the influence of the mass splitting on the first order electroweak phase transition, and find that it can further enhance the transition strength and thus induce gravitational waves during the phase transition, which may be detected in the forthcoming detectors such as U-DECIGO.
\end{abstract}


\section{Introduction}
With high precision, the Standard Model (SM) explains the interactions of all known fundamental particles. Despite of intensive investigations, no significant deviations from the SM have been identified in the experiments, suggesting that the SM constitutes the complete description of Nature. However, several long-standing issues imply that new physics beyond the SM is inevitable. The origin of neutrino masses and the formation of cold dark matter are the two primary puzzles that any extensions of the SM should address. Intriguingly, the newly measured $W$ boson mass provides yet another impetus for new physics. Very recently, with the improved parton distribution functions of the (anti)proton and new track reconstruction, the CDF collaboration at Fermilab has released the world's most precise direct measurement of the $W$ boson mass \cite{CDF:2022hxs}, $m_W^{\mathrm{CDF}} = 80.4435 \pm 0.0094~\mathrm{GeV}$, based on $8.8~\mathrm{fb}^{-1}$ of data gathered between $2002$ and $2011$, which is approximately $7\sigma$ away from the SM prediction $m_W^{\rm EW}=80.3545 \pm 0.0059$~GeV~\cite{deBlas:2021wap}. Furthermore, there is a significant tension between the new CDF result and the direct measurements implemented by the D0 collaboration at the Tevatron \cite{D0:2012kms} and the ATLAS/LHCb collaboration at the Large Hadron Collider (LHC) \cite{ATLAS:2017rzl,LHCb:2021bjt} while the latter ones are in good agreement with $m_W^{\rm EW}$. Although it is yet premature to confirm the existence of new physics, a number of analyses on the new data and interpretations in terms of physics beyond the SM have been carried out in the literature~\cite{Fan:2022dck, Lu:2022bgw, Athron:2022qpo, Yuan:2022cpw, Strumia:2022qkt, Yang:2022gvz, deBlas:2022hdk, Zhu:2022tpr, Addazi:2022fbj}.

In this work, we aim to bind the three issues together and interpret them in a single setup, where we investigate the scotogenic model consisting of three right-handed neutrinos and one $SU(2)$ doublet. This matter content was first postulated in \cite{Ma:2006km}, which uses the scotogenic scalar as dark matter, and explains the origin of tiny neutrino masses via a loop-induced process involving the new matter contents. We found that the mass splitting between the charged and neutral components of the $SU(2)$ doublet behaves like a bridge to connect the three issues naturally. The scotogenic model is one of the simplest loop realizations of the dim-5 Weinberg operator \cite{Weinberg:1979sa}. It lowers the seesaw scale from GUT to the electroweak (EW) scale, meanwhile provides candidate for dark matter in our Universe. The presence of a scotogenic scalar also explains the latest $W$ boson mass discrepancy. This is mainly because the authors in \cite{Lu:2022bgw} demonstrated that the oblique parameters $S$, $T$, and $U$ must deviate from the SM estimate to establish a consistent EW global fit, indicating the presence of mass-splitting in the scalar sector. Since the scotogenic scalar has an $SU(2)$ representation, it yields a non-vanishing contribution to oblique parameters.

The scalar sector in scotogenic model is also well known as the inert doublet  model~\cite{Deshpande:1977rw,LopezHonorez:2010tb,Arhrib:2012ia,Borah:2012pu,Gil:2012ya} which naturally includes a DM candidate, i.e. either the CP even Higgs or CP odd Higgs. However, the main difficulty in the original inert model is to produce the correct relic density of dark matter under the direct detection limits. Only in the Higgs funnel region (and heavy mass region), can inert scalar annihilation attain the proper relic density.  On the other hand, the right-handed neutrino in this model can become a promising dark matter candidate in most parameter space. The distinctive structure of Yukawa couplings is not only helpful to achieve tiny neutrino mass but beneficial to the freeze-in production of dark matter. The Feebly Interacting Massive Particle (FIMP) couples to the thermal bath extremely weak, so that it can not retain chemical equilibrium with the thermal bath. Instead, the DM particles were produced by the decay processes from the scalar sector, which provides the correct relic abundance. In addition, the new introduced particles could also be used to explain the anomalies in flavor physics \cite{Li:2018lxi}.

The predictability of this paradigm is an essential part of its attractiveness. Including the scotogenic scalar, the extra degrees of freedom lead to an inevitable first order electroweak phase transition (EWPT) in the early Universe. The scalar mass splittings required for explaining the $W$ boson mass happen to affect the magnitude of phase transition. EWPT in the scotogenic scalar sector has been extensively studied \cite{Borah:2012pu,Gil:2012ya,Blinov:2015vma,Fabian:2020hny}. However, the new $W$ boson mass anomaly strongly motivates us to re-study the brand new parameter space. As a consequence, the strong enough phase transitions assure the generation of gravitational waves (GWs). Astonishingly, we can use the resulting gravitational wave created to dig insight into the $W$ bosons with the future facilities. The essential characteristics and forecasts of the scotogenic model are presented in this work.
	
\section{Neutrino mass and dark matter}
The scotogenic model \cite{Ma:2006km} is one of the simplest ways to link neutrino mass to dark matter. Besides the particle content in SM, an additional doublet scalar (denoted as $H_2$) and three generations of right-handed neutrinos $N_k(k=1,2,3)$ are introduced. The new particles have odd parity under a new $Z_2$ symmetry, while the SM particles are all $Z_2$ even. With $Z_2$ odd nature, the right-handed neutrinos cannot form Dirac masses $\bar{L} \tilde H_1 N+\rm{h.c.}$, where $H_1$ represents the SM Higgs doublet and $\tilde H_1 \equiv \epsilon H_1^\ast$. The relevant terms in the Lagrangian, concerning right-handed neutrinos, are written as
\begin{equation}
    -\mathcal{L}_N = \frac{1}{2} m_N \overline{N^c} N + y_N \bar{L} \tilde H_2 N + \rm{h.c.}.
\end{equation}
Similar to the type-I seesaw, the right-handed neutrinos could have Majorana masses. The scalar potential, obeying SM gauge symmetries plus the discrete $Z_2$, is given as
\begin{eqnarray}
V= -\mu_1^2 |H_1|^2 + \mu_2^2 |H_2|^2 + \lambda_1 |H_1|^4
+ \lambda_2 |H_2|^4  +\lambda_3 |H_1|^2 |H_2|^2+ \lambda_4 |H_1^\dagger H_2|^2
+ \frac{\lambda_5}{2} \left\{ (H_1^\dagger H_2)^2 + \rm{h.c.} \right\}.
\label{potential}
\end{eqnarray}
The conserved $Z_2$ parity forbids $H_2$ from developing a vacuum expectation value after EW symmetry breaking. However, we should mention that the existence of an additional doublet alters the EWPT, which imprints the detection signature as gravitational waves.  The two doublets could be expressed as,
\begin{equation}
H_1 =
   \left( \begin{array}{c}  \phi_1^+ \\ \frac{1}{\sqrt{2}}(v + h + i \chi)  \end{array}
     \right),
H_2 =    \left( \begin{array}{c}  H^+ \\
\frac{1}{\sqrt{2}}(H  + i A)  \end{array}  \right),
\end{equation}
$h$ would be the only physical component of $H_1$, and it plays the role of the observed Higgs scalar. For $H_2$, the four components are all physical, with two neutral scalars, $H$(CP-even) and $A$(CP-odd), and two charged scalars, $H^\pm$. Generally, the quartic couplings can be written in terms of the physical scalars masses and $\mu_2$,
%
\begin{eqnarray}
\lambda_3 &=& \frac{2}{v^2}\left(m_{H^\pm}^2 - \mu_2^2\right), \label{eq:lambda3}\\
\lambda_4 &=& \frac{\left(m_H^2 + m_A^2 - 2 m_{H^\pm}^2\right)}{v^2},\label{eq:lambda4}\\
\lambda_5 &=& \frac{\left(m_H^2 - m_A^2\right)}{v^2}.\label{eq:lambda5}
\end{eqnarray}

We see that the $\lambda_5$ term controls the mass difference between the two neutral components, and together with $\lambda_4$, the mass differences in $H_2$ are determined. The couplings $\lambda_i$ are constrained by the unitarity, vacuum stability, and perturbativity requirements (see \cite{Arhrib:2012ia} for details), and in the following analysis we have taken these constraints into account.

Now one could give a calculation on the neutrino mass in the scotogenic model, which is given by \cite{Molinaro:2014lfa}
\begin{eqnarray}
	\left(\mathcal{M}_\nu\right)_{\alpha\beta}= \sum_{k=1}^3 \frac{y_{N}^{\alpha k} y_{N}^{\beta k}}{32\pi^2} m_{N_k}\Big[\frac{m_H^2}{m_H^2 - m_{N_k}^2} \ln \Big( \frac{m_H^2}{m_{N_k}^2}\Big)-\frac{m_A^2}{m_A^2 - m_{N_k}^2} \ln \Big( \frac{m_A^2}{m_{N_k}^2}\Big)\Big],
\end{eqnarray}
here $\alpha,\beta$ label the neutrino flavor indices. Thus the mass difference between $H$ and $A$ plays an important role to produce the loop neutrino mass, under the degenerate limit, i.e. $m_H =  m_A$(or $\lambda_5 = 0$),  the loop mass would be vanished. If $\lambda_5 \ll 1$, the mass matrix could be rewritten as
\begin{eqnarray}
    \left(\mathcal{M}_\nu\right)_{\alpha \beta}= \frac{\lambda_5 v^2}{32\pi^2}\sum_{k=1}^3 y_{N}^{\alpha k} y_{N}^{\beta k} \frac{m_{N_k}}{m_0^2 - m_{N_k}^2}
    \times \Big[1 - \frac{m_{N_k}^2}{m_0^2 - m_{N_k}^2} \ln \Big( \frac{m_0^2}{m_{N_k}^2}\Big)\Big],
\end{eqnarray}
with $m_0^2 = (m_H^2+m_A^2)/2$. The EW scale realization of neutrino mass at the present energy frontier is relevant to the low-energy experiment. Hence we focus on the masses of particles in loop below TeV. Apart from that, we would pay our attention to thermal freeze-in production of the dark matter, which is the lightest right-handed neutrino $N_1$. The out-of-equilibrium accumulation, which we discuss in the dark matter section, requires relevant coupling $y_N^{\alpha 1}$ to be extremely small. Hence $N_1$ contribution to neutrino mass is negligible. Such a scenario is still consistent with the neutrino oscillation result, since the converting probabilities care for the squared mass differences of neutrinos, while one vanished neutrino mass is allowed. One could estimate that to realize the neutrino mass at the TeV scale, a combining constraint is $\lambda_5 (y_N^{\alpha 2,3})^2 \sim 10^{-11}$ \cite{Molinaro:2014lfa}. It could be reached by tiny $\lambda_5$ while large $y_N^{\alpha2,3}$. But actually a too large Yukawa, e.g. $y_N>0.1$, would result in strong contradiction with the non-observation of lepton flavor violation(LFV) decays, e.g. $\ell_i \to \ell_j \gamma$.

The signature searching on collider of the new particles depends on relative sizes of their masses. If the charged scalar is the next-to-lightest $Z_2$ odd particle, it has no choice but to decay into $N_1$, leaving a highly ionizing charged track in the detector. The null result of searching for such kind of track from heavy stable charged particle, on LHC, could exclude $m_{H^\pm}$ up to $\sim 500~\rm{GeV}$ for $m_{N_1}\geq 1~\rm{MeV}$. If the next-to-lightest $Z_2$ odd particle is $N_2$, the searching signature would be two prompt charged leptons plus missing energy, which is from the decay of $H^\pm \to N_2 \ell^\pm$. The small decay rate of $N_2 \to \ell \bar{\ell} N_1$ would result in a decay-length which largely exceeds size of the detector. Searching for two prompt leptons plus missing energy on LHC could exclude $m_{H^\pm}$ up to $\sim 160~\rm{GeV}$, when assuming the prompt charged leptons are either $e^\pm$ or $\mu^\pm$. The exclusion limit would become weaker when decay into tau lepton is allowed. If $N_3$ is lighter than the charged scalar, the decay of $H^\pm\to N_3 \ell^\pm$ would open and $N_3$ could subsequently decay into $\ell \bar{\ell} N_2$. The relevant Yukawa couplings are strongly constrained by the non-observation of $\mu \to e \gamma$, hence may result in a signature of displaced lepton pair. The null result on LHC of the displaced lepton pair could set constraints on masses of $N_{2,3}$ and $H^\pm$, depends on the magnitude of Yukawa couplings. Decay products of the neutral scalars are similar to the charged one except that charged leptons are replaced by neutrinos, hence the limits on $m_{H/A}$ are weaker than the charged one. As we focus on the case that charged scalar is the lightest $Z_2$ odd scalar, adopting the above constraints for $H^\pm$ are enough for the scalar sector. For detailed discussion of the collider searching, one is referred to \cite{Hessler:2016kwm}. In the following discussions, we will adopt the corresponding constraints on masses whenever the mass spectra are fixed. Besides, requirements on neutrino mass and obeying LFV decays constraints are also imposed.

The discrete $Z_2$ parity on one hand gives loop masses for neutrinos, on the other hand, it also suggests a candidate particle for dark matter. In principle, both the lightest neutral scalar and sterile neutrino can be dark matter. For the former situation, the neutral scalar could be thermally freeze-out produced and becomes a typical WIMP. However, due to the current stringent limits from the direct detection and relic density, most of the parameter space has been ruled out of the WIMPs\cite{Fabian:2020hny}. Another possibility is that the lightest right-handed neutrino, i.e. $N_1$, acts as a FIMP dark matter. In this situation, dark matter $N_1$  is produced out of thermal equilibrium. Specifically speaking, $N_1$ is accumulated from the decay of $ X(H, A, H^{\pm}) \to N_1 L$, which relates to Yukawa $y_N^{\alpha 1}$ ($y_1$ for short thereafter). The out of equilibrium condition requires the decay rate to be smaller than the rate of Universe expansion, i.e. $\Gamma(X \to N_1 L) < H(T\sim m_{X})$, which would force $y_1$ to be small. The dark matter $N_1$ yield, $Y_{N_{1}}(T)=n_{N_1}(T)/s(T)$, can be computed by solving  the following Boltzmann equation \cite{Hall:2009bx}
\begin{equation}
	sT\frac{d Y_{N_1}}{dT} = - \frac{\gamma_{N_1}(T)}{H(T)},
\end{equation}
with $\gamma_{N_1}(T)$ representing the thermal averaged FIMP production rate
\begin{equation}
	\gamma_{N_1}(T) = \sum_X \frac{g_X m_X^2 T}{2\pi^2} K_1(m_X/T) \Gamma(X\to N_1 \ell),
\end{equation}
$g_X$ is the internal degrees of freedom of $X$,
$s$ is the entropy density of the Universe, $H(T)$ is the expansion rate of the Universe at a given temperature and  $K_{1}(x)$ is the Bessel function of the second kind.

The relevant study can be found in Ref.~\cite{Molinaro:2014lfa}. $N_1$ production will be dominated by the decays of the scalars ($H,A,H^\pm$) while they are in equilibrium with the thermal bath. The accumulation of $N_1$ could, in principle, also from decay of heavier sterile neutrinos, i.e. $N_{2,3} \to N_1 \bar{\ell} \ell$, but Ref.~\cite{Molinaro:2014lfa} has verified these decays are subdominate.

The decay rates that enter into $\Gamma\left(X\to N_{1} \, L\right)$ are calculated as,
\begin{align}\label{decaytoN1}
\Gamma(H\to N_1 \bar{\nu} )&=\frac{(m_H^2-m_{N_1}^2)^2 }{32\pi m_H^3}y^2_{1},\\
\Gamma( A\to N_1\,\bar{\nu})&=\frac{(m_A^2-m_{N_1}^2)^2  }{32\pi m_A^3}
y^2_{1},\\
\Gamma( H^+\to N_1\,\bar{\ell})&=\frac{(m_{H^+}^2-m_{N_1}^2)^2 }{32\pi m_{H^+}^3}y^2_{1}.
\end{align}
By simply numerical calculations we can obtain the observed DM relic density with the Yukawa coupling in the range $y_1\in [10^{-12},  10^{-11}$], for tens to hundreds GeV FIMPs. To study the correlations with the new CDF measured $m_W$ and EWPT, we take $m_{H^{\pm}}=300$ GeV, $y_1=1.48\times10^{-12}$ and $m_{N_1}=100$ GeV as a benchmark. Then we find the viable parameter space for generating the relic density $\Omega h^2=0.12\pm0.0012$ \cite{Planck:2018vyg, Abdughani:2021pdc}, which is shown by the purple band on mass splittings plane, see Fig.~\ref{fig:all}.

\section{$W$ boson mass and Gravitational Waves}

The relationship between $W$ boson mass and the oblique corrections is given by \cite{Peskin:1991sw},
\begin{eqnarray}
\label{eq:WST}
m_{W}^{2}=m_{W}^{2}(\mathrm{SM})+\frac{\alpha c^{2}}{c^{2}-s^{2}} m_{Z}^{2} [-\frac{1}{2} \Delta S +c^{2} \Delta T +\frac{c^{2}-s^{2}}{4 s^{2}} \Delta U ]\, ,
\end{eqnarray}
where $c=\cos\theta_W$, $s=\sin\theta_W$.
To generate suitable $W$ boson mass, we resort to a significant contribution from $\Delta S$ and $\Delta T$.
The expressions are  \cite{Arhrib:2012ia}
\begin{eqnarray}
\Delta S= \frac{1}{2\pi}\Big[ \frac{1}{6}\log(\frac{m^2_{H}}{m^2_{H^\pm}}) -
  \frac{5}{36} + \frac{m^2_{H} m^2_{A}}{3(m^2_{A}-m^2_{H})^2} + \frac{m^4_A (m^2_{A}-3m^2_{H})}{6(m^2_{A}-m^2_{H})^3} \log(\frac{m^2_{A}}{m^2_{H}})\Big],
\end{eqnarray}
and
\begin{eqnarray}
\Delta T = \frac{1}{32\pi^2 \alpha v^2}\Big[ F(m^2_{H^\pm}, m^2_{A})
+ F(m^2_{H^\pm}, m^2_{H})
- F(m^2_{A}, m^2_{H})\Big]
\end{eqnarray}
where the function $F$ is defined by
\begin{equation}
	F(x,y) = \left\{ \begin{array}{lr}
		\frac{x+y}{2} - \frac{xy}{x-y}\log(\frac{x}{y}), & x\neq y,  \\
		0, & \, x=y.
		\end{array}  \right.
\end{equation}

\begin{figure}[!htbp]
	\centering
	\includegraphics[width=0.65\textwidth]{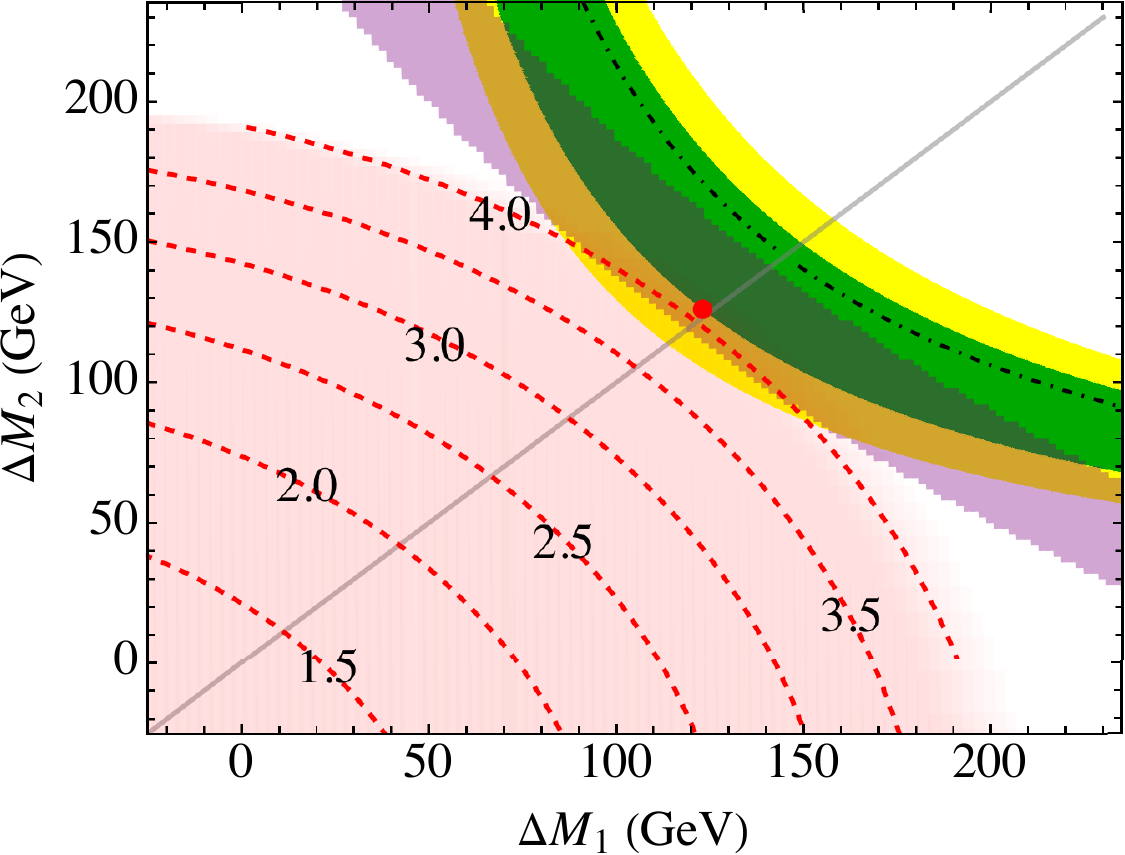}
	\caption{(Color online) Best fit region for $\Delta S, \Delta T$ operators, FOPT strength $v_n/T_n$ and DM relic density,  as functions of $\Delta M_1$ and $\Delta M_2$ where $\Delta M_1\equiv m_H-m_{H^\pm}$ and  $\Delta M_2\equiv m_A-m_{H^\pm}$.
	The green and yellow bands correspond to the best fit with 1- and 2-sigma regions required by the new CDF measured $m_W$.
	The black dash-dotted line satisfies the central value of the CDF $W$ mass.
	In the red region the strongly FOPT can be realized and the red dashed lines represent the contours of $v_n/T_n$.
	The gray line represents the condition $m_H=m_A$ which means that one cannot obtain the neutrino mass here.  While the neutrino mass can be generated in the most parameter space with adjusted Yukawa couplings.
	In the purple region, one can yield the correct DM relic density $\Omega h^2=0.12\pm0.0012$ \cite{Planck:2018vyg},
	with coupling $y_1=1.48\times10^{-12}$ and $m_{N_1}=100$ GeV.
    The red point is the selected benchmark point (BP) to investigate GWs physics below, which is: $m_H=423$ GeV, $m_A=426$ GeV, $m_{H^\pm}=300$ GeV, $\lambda_3=1.5$, $\lambda_4=2.98$, $\lambda_5=-0.04$, $T_n=158.78$ GeV, $v_n/T_n=4.1$.
	}
	\label{fig:all}
	\end{figure}

The $T$ parameter vanishes when $m_{H^{\pm}}=m_H$ or $m_{H^{\pm}}=m_A$, since these conditions lead to an exact custodial $SU(2)$ symmetry, in which one of the neutral scalars joins the charged scalars to create a $SU(2)$ triplet. So to get proper oblique corrections, one knows that the key characters are the mass splittings of the $Z_2$ odd scalars, namely $\Delta M_1\equiv m_H-m_{H^\pm}$ and  $\Delta M_2\equiv m_A-m_{H^\pm}$. In previous studies of scalar multiplets models, the mass splittings are limited by the previous $S,~T$ parameters fitted by the $m_W$ mass of PDG. Now we have the right to reexamine the whole picture.  In terms of $\Delta M_1$ and $\Delta M_2$ we could directly get the CDF $W$ mass (the central value) by using Eq.~\ref{eq:WST}, which is shown in Fig.~\ref{fig:all} by the black dash-dotted line in $\Delta M_1$ and $\Delta M_2$ plane. We take $m_{H^\pm}=300$ GeV as a benchmark model. In order to find the viable parameter regions, one should also consider the correlations to the $W$ mass from the EW parameters. So we use the global EW best fit values of $\Delta S$ and $\Delta T$ derived in Ref.~\cite{Lu:2022bgw} (Table-III) to perform a fit by using  $\chi^2({\bf O})=({\bf y}-\mu({\bf O}))^T {\bf C}^{-1}({\bf y}-\mu({\bf O}))$, where ${\bf y}$ is the vector of central values and ${\bf C}$ is the covariance matrix.  In Fig.~\ref{fig:all} we show 1-and 2-sigma regions allowed by $\Delta S$ and $\Delta T$ parameters in the mass splittings plane. The $\Delta T$ parameter becomes negative if the charged Higgs mass $m_{H^{\pm}}$ falls into the range between the masses of the two neutral scalars, $m_A$ and $m_H$. As a result, either $m_{H^{\pm}} > m_{A,H}$ or $m_{H^{\pm}} < m_{A,H}$ are permitted. Because of $S \sim  \log(m_{H,A}/m_{H^{\pm}})$, the $S$ parameter is disposed to be negative in the former scenario, which is disfavored by the global fit \cite{Lu:2022bgw}. Then we only consider the case that charged Higgs is the lightest one in new higgs doublet.

The rich spectrum of the scalar sector also drives the evolution dynamics of the vacuum state non-trivial and provides the first order EWPT in the early Universe. By considering the thermal loop effects we use the approximated effective potential near the critical temperature \cite{Chung:2012vg}
\begin{align}\label{vappro}
	V_{\rm eff}(h, T) \approx \frac{1}{2} ( -\mu^2 + c T^2) h^2 - \frac{ \varepsilon T}{12 \pi} (h^2)^\frac{3}{2} + \frac{\lambda}{4} h^4,
\end{align}
where the coefficient $\varepsilon$ quantifies the interactions between the extra scalars and the Higgs boson. We have
$\varepsilon \approx (6 m_W^3 + 3 m_Z^3)/{v^3} + 2 (\lambda_3/{2})^{3/2} +((\lambda_3+\lambda_4 - \lambda_5)/{2})^{3/2} + ((\lambda_3 + \lambda_4 + \lambda_5)/2)^{3/2}$
and
$c \approx (6 m_t^2+6 m_W^2+3m_Z^2+\frac{3}{2}m_H^2)/(12 v^2)+ (2\lambda_3+\lambda_4)/12$.

The sizable oblique corrections favor large mass splittings $\Delta M_1$ and $\Delta M_2$ in order to interpret the new measured $W$ boson mass, as shown in Fig.~\ref{fig:all}. So the quartic couplings $\lambda_4, \lambda_5$ in the effective potential directly link the new $W$ deviation and the EWPT, then  lead to an affected EWPT.

The symmetry breaking phase starts to nucleate right after the Universe cooling to the nucleation temperature $T_n$. The nucleation temperature is defined by the equality of the nucleation rate per Hubble volume and the Universe expansion rate, i.e. $\Gamma(T_n)=H^4(T_n)$, where $\Gamma(T)\sim T^4e^{-S_3/T}$ is the decay rate per unit volume and  $S_3$ is the classical action for the $O(3)$ symmetric bounce solution~\cite{Linde:1981zj}. For a radiation-dominated Universe and a first order phase transition (FOPT) happening around the EW scale, $T_n$ can be solved by $S_3/T_n\sim140$~\cite{Quiros:1999jp}. This criterion will be taken as the sufficient condition for a FOPT in this work.  Further we use the package {\tt cosmoTransition}~\cite{Wainwright:2011kj} to calculate the bounce solution and $T_n$ for the  effective potential $V_{\rm eff}$. The custodial symmetry in the scalar potential leads to a mass degeneration of $m_A$ and $m_{H^{\pm}}$ in previous research (or before the $W$ mass deviation is identified).  Generally, the EWPT strength depends on the mass splitting since it affects the potential barrier from custodial symmetry breaking. Now we have the chance to open the new parameter space by taking  $m_A \neq m_{H^\pm}$, we can find that such mass splitting $\Delta M_2$  enhances the EWPT strength in an obvious way, as shown in Fig.~\ref{fig:all} with red-dashed contours. The reason is mainly that the larger $\Delta M_1$ and/or $\Delta M_2$ leads to a higher barrier between the symmetric vacuum and the symmetry breaking one.  (see Eqs.~\ref{eq:lambda4}, \ref{eq:lambda5} and Eq.~\ref{vappro}). This conclusion could also be generalized to other scalar multiplets models.

During a FOPT, stochastic GWs come from three sources: bubble collisions \cite{Huber:2008hg,Di:2020ivg}, sound waves in the plasma~\cite{Hindmarsh:2015qta,Guo:2020grp} and the magneto-hydrodynamics turbulence~\cite{Binetruy:2012ze,Caprini:2009yp}. Recent studies show that the bubble collision contribution to the GWs can generally be neglected because only a tiny fraction of the FOPT energy deposits in the bubble wall\cite{Bodeker:2017cim}. It turns out that the dominant contribution comes from the sound waves as most FOPT energy is pumped into the surrounding fluid shells~\cite{Ellis:2018mja}. Then the turbulence is another main source after a finite period of the sound wave\cite{Ellis:2020awk}. Consequently, the GWs spectrum today can be expressed as (see details in supplementary file.)
\begin{equation}
\Omega_{\rm GW}(f)=\Omega_{\rm sw}(f)+\Omega_{\rm turb}(f),
\end{equation}
where $f$ is the frequency, the subscripts ``sw'' and ``turb'' denote sound waves and turbulence respectively.

\begin{figure}[!htbp]
	\centering
	\includegraphics[width=0.65\textwidth]{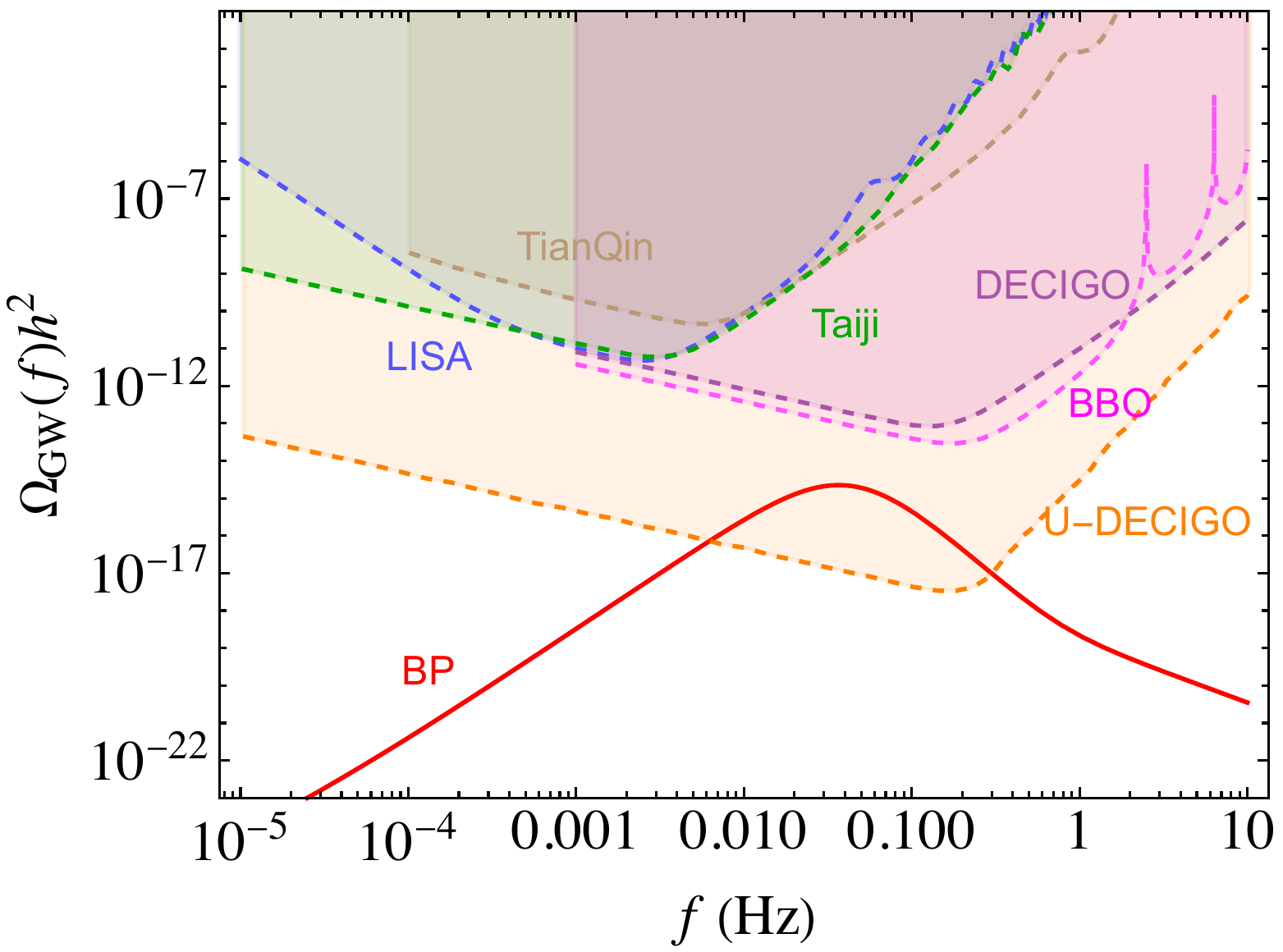}
	\caption{(Color online) GW spectrum $\Omega_{\rm GW} h^2$ from the benchmark point where $\alpha=0.021$, ${\beta}/{H}=722.9$, depicted with the red line. The colored regions represent the expected sensitivities of the projected GWs interferometers.
	}
	\label{fig:gw}
	\end{figure}

The GWs produced from a FOPT around EW scale have the potential to be probed by the next generation space-based laser interferometers such as
LISA~\cite{LISA:2017pwj},
BBO~\cite{Crowder:2005nr},
TianQin~\cite{TianQin:2015yph, Hu:2017yoc},
Taiji~\cite{Hu:2017mde, Ruan:2018tsw, Zhao:2019gyk},
DECIGO~\cite{Kawamura:2011zz,Kawamura:2006up} and
U-DECIGO~\cite{Kudoh:2005as}.
In Fig.~\ref{fig:gw} we plot the GW spectrum with the benchmark point. We find that the GW signals are within the detectability of U-DECIGO,  and this possibility gives a complementary detection to the other searches. While to produce stronger signals one may consider the extension of the scalar sector, which further enhances the sensitivities ~\cite{Bian:2018bxr}.

\bigskip
	
\section{Conclusion}
Eventually, we successfully explain our Universe with the new measured $W$ boson mass, the correct abundance of FIMP dark matter, and the Majorana neutrino masses within a minimal extension of the SM. We found that both neutrino mass and $W$ boson mass are sensitive to the non-vanishing $\Delta M_1$ and $\Delta M_2$, which also influences the freeze-in production of sterile neutrino dark matter. As a result, the non-degeneration between the charged and neutral components of a scotogenic scalar naturally links the three fundamental issues. More importantly, this model makes a precise prediction that will be validated in future GW experiments. The additional multiplet in the scalar sector drives the evolution dynamics of the vacuum state non-trivial. Thus in the early Universe, it provides a first-order electroweak phase transition, which is further enhanced by the new required mass splitting $m_A-m_{H^\pm}$. This is thought to be an efficient approach to generating detectable GWs. The future gravitational wave detectors will help reveal valuable information about the nature of this model.


\section*{Acknowledgements}
This work was supported by the National Natural Science Foundation of China under the grants Nos. 11805161, 12005180, and  11975195, by the Natural Science Foundation of Shandong Province under Grant No. ZR2020QA083 and ZR2019JQ04 and by the Project of Shandong Province Higher Educational Science and Technology Program under Grants No. 2019KJJ007.


\bibliographystyle{bibstyle}
\bibliography{biblist}
\end{document}